\documentclass[twocolumn,aps,prd,showpacs,nofootinbib,superscriptaddress,preprintnumbers,floatfix,10pt]{revtex4-1}

\usepackage{graphicx}  
\usepackage{bm}        
\usepackage{amssymb}   
\usepackage{hyperref}
\usepackage[english]{babel}

\usepackage{color}

 \usepackage{longtable}
 \usepackage{amsmath}
 \usepackage{feynmf}

\newcommand*{\no}{\noindent}
\newcommand*{\bea}{\begin{eqnarray}}
\newcommand*{\eea}{\end{eqnarray}}
\newcommand*{\be}{\begin{equation}}
\newcommand*{\ee}{\end{equation}}

\newcommand*{\pref}[1]{(\ref{#1})}

\newcommand*{\prefr}[2]{(\ref{#1}-\ref{#2})} 
\newcommand*{\nn}{\nonumber}

\newcommand*{\bma}{\begin{pmatrix}}
\newcommand*{\ema}{\end{pmatrix}}

\newcommand*{\la}{\langle}
\newcommand*{\ra}{\rangle}

\begin{document}

\preprint{}

\title{Pair production processes and flavor in gauge-invariant perturbation theory}

\author{Larissa Egger}
\email{larissa.egger@uni-graz.at}
\affiliation{Institute of Physics, NAWI Graz, University of Graz, Universit\"atsplatz 5, A-8010 Graz, Austria}
\author{Axel Maas}
\email{axel.maas@uni-graz.at}
\affiliation{Institute of Physics, NAWI Graz, University of Graz, Universit\"atsplatz 5, A-8010 Graz, Austria}
\author{Ren\'e Sondenheimer}
\email{rene.sondenheimer@uni-jena.de}
\affiliation{Institute for Theoretical Physics, Friedrich-Schiller-University Jena, Max-Wien-Platz 1, D-07743 Jena, Germany}

\begin{abstract}
Gauge-invariant perturbation theory is an extension of ordinary perturbation theory which describes strictly gauge-invariant states in theories with a Brout-Englert-Higgs effect. Such gauge-invariant states are composite operators which have necessarily only global quantum numbers. As a consequence, flavor is exchanged for custodial quantum numbers in the standard model, recreating the fermion spectrum in the process. Here, we study the implications of such a description, possibly also for the generation structure of the standard model.

In particular, this implies that scattering processes are  essentially bound-state-bound-state interactions, and require a suitable description. We analyze the implications for the pair-production process $e^+e^-\to{\bar f}f$ at a linear collider to leading order. We show how ordinary perturbation theory is recovered as the leading contribution. Using a PDF-type language, we also assess the impact of sub-leading contributions. To lowest order we find that the result is mainly influenced by how large the contribution of the Higgs at large $x$ is.
This gives an interesting, possibly experimentally testable, scenario for the formal field theory underlying the electroweak sector of the standard model.
\end{abstract}

\maketitle

\section{Introduction}

Gauge invariance of experimental observables is a fundamental requirement of theories like the standard model \cite{'tHooft:1979bj,Banks:1979fi,Frohlich:1980gj,Maas:2012tj}. In the electroweak sector, this leads to an apparent contradiction. Strictly speaking, the elementary particles, i.e., the fields of the Lagrangian, the Higgs, the gauge bosons, but also the fermions, are not gauge-invariant states \cite{'tHooft:1979bj,Banks:1979fi,Frohlich:1980gj}. However, treating them like they would be in perturbation theory gives an excellent description of experiments \cite{pdg,Bohm:2001yx}.

This paradox is resolved by the Fr\"ohlich-Morchio-Strocchi (FMS) mechanism \cite{Frohlich:1980gj,Frohlich:1981yi}. This mechanism yields that to a very good approximation the properties of gauge-invariant, and thus composite, bound states coincide with those of the perturbative ones, at least in gauges in which perturbation theory is applicable \cite{Lee:1974zg}. This effect has also been confirmed in lattice simulations for the static properties of the particles \cite{Maas:2012tj,Maas:2013aia}.

However, this poses the question of what are the implications for dynamics, especially the scattering of particles. The basic recipe of the FMS mechanism \cite{Frohlich:1980gj,Frohlich:1981yi} can in principle be extended also to this situation \cite{Maas:2012tj,Torek:2016ede}. This describes how the scattering of two bound states is to leading order given by the scattering process of the (leading) elementary constituents \cite{Frohlich:1980gj,Frohlich:1981yi,Maas:2012tj}, a picture not dissimilar to the scattering of QCD bound states \cite{Bohm:2001yx}. There is, however, one important difference. In the standard picture of the electroweak sector only one of the actual constituents of the bound state operators contributes to all orders in conventional perturbation theory. Capturing the contribution from the second part requires to go beyond conventional perturbation theory in the FMS mechanism.

Doing so is the aim of this work. The main motivation is not that large deviations from the standard model are necessarily expected. In fact, as the following will show, deviations are probably restricted to very special circumstances, if at all. This is likely due to the particular structure of the standard model. In theories beyond the standard model this may change \cite{Maas:2015gma,Torek:2016ede,Maas:2014pba,Maas:2016ngo}, though this does not need to be necessarily so \cite{Maas:2016qpu}. Nonetheless, understanding the standard model case, and at least identifying possible candidate scenarios where a difference could be expected, is a necessary first step. This will be done here.

To this end, we will investigate pair production in the process $e^+e^-\to{\bar f}f$, the situation at LEP(2) and the planned ILC and CEPC. According to the FMS mechanism, a physical electron is actually a gauge-invariant bound state formed by the elementary electron and the elementary Higgs, see section \ref{s:flavor}. To distinguish such bound states from the elementary ones, we will denote them by capital letters, both for the symbols and the name. Likewise, the elementary states will receive for names and symbols small letters. For instance, we will use $H$ for the bound state Higgs while $h$ denotes the elementary field of the Lagrangian. Thus, the actually investigated process will be $E^+E^-\to\bar{F}F$.

As will be seen, any effects are probably not to be expected below the 2-higgs threshold, making this CEPC and ILC physics.

However, in a gauge-invariant setting already this result requires some amendments to the usual perturbative setup. Why this is the case, and what kind of far-reaching implications it has, will be discussed in section \ref{s:flavor}. The actual framework for the calculation of the process will be developed in section \ref{s:gipt}. This leads to a description\footnote{Interestingly, an approach based on a confining rather than a Brout-Englert-Higgs-type physics in the standard model leads to formally quite similar results \cite{Calmet:2000th,Calmet:2001rp,Calmet:2001yd,Calmet:2002mf}, though for completely different physics reasons. Similar lines of arguments \cite{Dosch:1983hr,Dosch:1984ec} also follow in the Abbott-Farhi model \cite{Abbott:1981re}, but again for different physics reasons.} quite similar to the one for bound state scattering in QCD using parton distribution functions (PDFs) \cite{Bohm:2001yx}. We apply this formalism to the aforementioned process in section \ref{s:dy}. We find that some deviations may be observable, depending on the actual structure of the bound states. We summarize the presentation in section \ref{s:sum}.

\section{Flavor}\label{s:flavor}

The basic idea behind the FMS mechanism is to formulate every observable first in a strictly, i.e., also non-perturbatively, gauge-invariant form. This is a far stronger statement than the usual perturbative gauge-invariance, which holds only in a limited class of gauges \cite{Lee:1974zg,Nielsen:1975fs}. In particular, in a non-Abelian gauge theory, like the electroweak sector\footnote{Note that such a theory always has an intact gauge symmetry \cite{Elitzur:1975im}. In the special case of the standard model, it is also, strictly speaking, not possible to distinguish confinement-type and Brout-Englert-Higgs-type physics \cite{Osterwalder:1977pc,Fradkin:1978dv,Caudy:2007sf,Seiler:2015rwa}.}, no single-field operator is gauge-invariant, and it is necessary to resort to composite operators.

However, composite operators are essentially bound state operators, and thus such states must be considered bound states. This has been investigated and supported in lattice calculations for the Gauge bosons and the Higgs boson \cite{Maas:2012tj,Maas:2013aia,Torek:2016ede}. In these references also details for these states, like the particular form of the corresponding composite operators, can be found. A brief review of the bosonic sector is given in   \cite{Torek:2016ede}. Here, the subject is different: the fermions.

\subsection{Custodial symmetry replaces flavor symmetry}\label{s:custodial}

Given a (left-handed) fermion in some fixed generation, its flavor is actually the weak gauge charge \cite{Bohm:2001yx}. Thus, the fermion state itself cannot be a gauge-invariant state \cite{Frohlich:1980gj,Frohlich:1981yi}.

It is therefore necessary to construct a state which can emulate the elementary fermions \cite{Frohlich:1980gj,Frohlich:1981yi}. As will be seen, a suitable gauge-invariant state is
\begin{align}
 \mathcal{O}^{g}(x) = h^{\dagger}(x) f^g(x), \quad \text{with} \qquad h = \begin{pmatrix} \phi_2^* & \phi_1 \\ -\phi_1^* & \phi_2 \end{pmatrix}.
 \label{fop}
\end{align}
\no The fermions are encoded in the field $f$, which is the (left-handed) weak doublet of generation $g$. The elements $\phi_i$ are the components of the usual higgs doublet. The hypercharge is skipped here, but can be added in a straightforward way \cite{Frohlich:1980gj,Frohlich:1981yi}. The matrix-valued higgs field $h$ transforms  under custodial transformations by multiplication from the right and under gauge transformations by multiplication from the left \cite{Shifman:2012zz}. Thus, this state is gauge-invariant, but remains a custodial doublet. Therefore, it has no longer any weak charge, but it inherits from the higgs field the feature to be a custodial (left-) doublet. Thus, the state contains two particles, distinguished by their global custodial charge. The Yukawa interaction eventually breaks the custodial symmetry, and creates the difference between the two states. Finally, the index $g$ separates generations, and also leptons and quarks. Thus, the identity of being a quark or a lepton, as well as the generation, is carried over from the fermion to the full state. Since the higgs is a scalar, spin and parity are also inherited from the fermion, and the operator on the left-hand side is a spinor. Thus, flavor doublets become custodial doublets.

Consider now, e.g., an electron and neutrino. The new gauge-invariant state is \cite{Frohlich:1980gj,Frohlich:1981yi}\footnote{Note that there is no distinct capital $\nu$, and thus $N$ is used.}
\begin{align}
 \mathcal{O}^{\nu e} = h^\dagger \begin{pmatrix} \nu \\ e \end{pmatrix} = \begin{pmatrix} \phi_2 \nu  -  \phi_1 e  \\  \phi_1^* \nu + \phi_2^* e  \end{pmatrix} \equiv \begin{pmatrix} N \\ E \end{pmatrix}
\end{align}
\no showing that this state actually mixes what is conventionally thought of as neutrino and electron.

The ordinary states do reemerge when applying the FMS mechanism \cite{Frohlich:1980gj,Frohlich:1981yi}. This requires to fix a gauge in which the higgs vacuum expectation value is non-zero, in the present work we use a 't Hooft gauge in which $\la\phi_i\ra=v\delta_{2i}$ holds. Rewriting $\phi_i=v \delta_{2i}+\varphi_i$ in ${\cal O}^{\nu e}$ leads in leading order in the fluctuation fields $\varphi$ to 
\be
{\mathcal O}^{\nu e}=v\bma \nu\cr e\ema+{\cal O}(\varphi)\label{fmssingle},
\ee
\no and by this the usual elementary doublet reemerges. In particular, any two-point function constructed from the operator \pref{fop} will therefore have the same mass poles, to this order in $\varphi$, as the elementary fields\footnote{For the higgs and the weak gauge bosons this mechanism has been demonstrated in lattice calculations to be working \cite{Maas:2012tj,Maas:2013aia}. This has also been seen for a toy GUT theory \cite{Maas:2016ngo}.}.

This also implies that the generation of the Dirac mass term for the matter particles becomes an even more subtle process and the particles themselves are rather sophisticated objects. 
From the gauge-invariant perspective (regarding the weak isospin), a 'physical' Dirac fermion $F$ consists now of a 'physical' left-handed fermion $F_L$, the propagation of which is described by a bound state of the elementary fermion $f_L$ and the Higgs field $\phi$, and a right-handed fermion $f_R\equiv F_R$ which is not charged under the weak isospin. In this way the elementary fields make up the propagator. The mass term mixes these two chirality states as usual but by a nontrivial interplay between the right-handed fermion and the left-handed fermion-higgs bound state via the Yukawa interaction.

Roughly speaking, the following picture emerges: Conventionally, and in a fixed gauge, an electron can start as a left-handed electron which flips to a right-handed one through an interaction with the higgs condensate. This yields to leading order the tree-level mass, and which can be resummed to include quantum corrections. The picture in a gauge-invariant description is quite different. A left-handed electron-higgs bound state can transform by a Yukawa interaction of its constituents into a right-handed electron and back. In fact, this is an oscillation phenomenon. This kind of dressing leads to the mass of the state. Thus, the Yukawa interaction is still responsible to create the masses of the fermions. But it is now a dynamical effect without the need of an explicit nonvanishing higgs condensate. A similar picture would arise in any fixed gauge with vanishing higgs vacuum expectation value \cite{Maas:2012ct}.  Even though at first a quite different picture, the FMS expansion connects them. The conventional picture reemerges as the leading-order contribution in an expansion in the fluctuation fields.

It is useful to also consider the full expression for each of the two custodial charges separately
\begin{align}
 \la \mathcal{O}^{\nu e}_1(x) \bar{\mathcal{O}}^{\nu e}_1(y) \ra 
 &= v^2 \la \nu \bar\nu \ra  + v \la (\varphi_2(x)+\varphi_2^*(y))\nu\bar\nu \ra \notag \\
 &\quad - v \la \varphi_1 e \bar\nu \ra  -  v \la \varphi_1^* e \bar\nu \ra  +  \la \varphi_2\varphi_2^*\nu\bar{\nu} \ra \notag \\
 &\quad +  \la \varphi_1\varphi_1^*e\bar{e} \ra  -  \la \varphi_1\varphi_2^* e\bar\nu \ra  -  \la \varphi_2\varphi_1^* \nu\bar{e} \ra,
 \\
 \la \mathcal{O}^{\nu e}_2(x) \bar{\mathcal{O}}^{\nu e}_2(y) \ra 
 &= v^2 \la e \bar{e} \ra  + v \la (\varphi_2^*(x)+\varphi_2(y))e\bar{e} \ra \notag \\
 &\quad + v \la \varphi_1 e \bar\nu \ra  +  v \la \varphi_1^* e \bar\nu \ra  +  \la \varphi_2^*\varphi_2e\bar{e} \ra \notag \\
 &\quad +  \la \varphi_1^*\varphi_1\nu\bar{\nu} \ra  + \la \varphi_2^*\varphi_1 e\bar\nu \ra  +  \la \varphi_1^*\varphi_2 \nu\bar{e} \ra,
\end{align}
\no where the arguments have only been indicated where they are not obvious. The two-point correlation functions describe the motion of the elementary particles through the condensate. This is just the usual picture in a fixed gauge. The four-point functions describe interactions of the constituents in this fixed gauge. The appearance of the other fermion field is possible, as these changes are balanced by the corresponding changes in the fluctuation fields. The three-point functions are special. They correspond to the absorption or emission of a fluctuation field in the final or initial state from a state, which previously or afterwards interacted with the condensate. This condensate interaction is necessary to balance the energy and custodial quantum number. Note that only the sum of all correlation functions is exactly gauge-invariant. So far, these correlation functions are the full correlation functions, in particular no perturbative expansion of them has been performed.

Consider now as leading order the zero coupling limit. This leads to a vanishing of the three-point functions. The two-point function remains, yielding the tree-level propagator for an electron and a neutrino. This gives the single particle pole of the FMS mechanism as discussed after Eq.~\pref{fmssingle}.

The four-point functions expand to products of two tree-level propagators, the corresponding fermion one and some of the higgs degrees of freedom. They correspond to a non-interacting two-particle state of a fermion, and a higgs or would-be goldstone. Those contributions involving the goldstones are BRST non-singlets, which will cancel. This leaves only BRST singlets, in this case the usual elementary higgs particle. This can be interpreted as a propagating fermion, which is accompanied by a higgs excited from the condensate. This therefore predicts a two-particle scattering pole on the left-hand side at the sum of the masses of the fermion and the higgs\footnote{In the purely bosonic correlators the corresponding scattering poles have indeed been seen on the lattice \cite{Maas:2014pba}.}.

At higher orders in conventional perturbation theory also connected three-point functions, indicating a scattering with the condensate, and connected four-point functions, initiating the scattering with an excitation from the condensate, contribute. However, such contributions are usually neglected when determining the propagation of elementary particles. Given that they are proportional to the fermion-higgs-Yukawa coupling, they should indeed be negligible for anything but for the top and, perhaps, the bottom. We will return to this in section \ref{s:gipt}.

This combination of gauge-invariance, the FMS mechanism, and conventional perturbation theory has been dubbed gauge-invariant perturbation theory \cite{Seiler:2015rwa}, and we will stick to this term\footnote{There is also another perturbative approximation with this name for the electroweak sector alone \cite{Langguth:1985eu,Philipsen:1996af}, which is actually applied to a slightly deformed theory \cite{Maas:2013aia}. However, it cannot be applied in the presence of QED or Yukawa couplings to fermions, and should not be confused with the present one.}.

There are two more remarks to be made.

The first is the custodial symmetry breaking combination $\la {\cal O}^{\nu e}_1(x)\bar{{\cal O}}_2^{\nu e}(y)\ra$. As the Yukawa couplings are non-zero, this will be non-zero. However, its leading order is $\la\nu \bar{e}\ra$, which is zero due to electric charge conservation. Thus in the operator $\la {\cal O}_1^{\nu e}\bar{{\cal O}}_2^{\nu e}\ra$ contributions arise only beyond two-point level. This is a rather generic feature of the FMS mechanism: If there is no elementary state with the corresponding quantum numbers, no identification with an elementary particle can be performed. However, correspondence can be here seen in a very broad sense \cite{Maas:2015gma,Torek:2016ede,Maas:2016ngo}, as the transformation from flavor to custodial quantum numbers already demonstrates.

The second issue arises from the other gauge interactions. As already noted, the Abelian nature of the hypercharge avoids any problem, as here a Dirac phase factor is sufficient \cite{Haag:1992hx}. Thus, the hypercharge will be ignored in the remainder of this section.

The strong interaction and quarks are different. At first it seems that the problem is irrelevant, as they are anyway bound in hadrons. This might be true for some meson states, as there is the possibility that all gauge quantum numbers can be contracted to a singlet. For instance, the state $(\bar{u}u + \bar{d}d)$, the $\omega$ Meson, is indeed invariant under weak $\mathrm{SU}(2)$ as well as the strong gauge group $\mathrm{SU}(3)_{\mathrm{c}}$. However, this will be obviously not the case for all Mesons, e.g., open-flavor Mesons. An example for this latter case are the charged Pions.

It is again necessary to exchange flavor for the custodial symmetry. This can also be formulated in a gauge-invariant manner regarding the weak isospin with the aid of the higgs field, analogously to the Lepton case, see Eq.~\eqref{fop}. Thus, e.g., the $\Pi^+$ can be described fully gauge-invariantly by the custodial state $\bar{\mathcal{O}}^{ud}_2\mathcal{O}^{ud}_1$ which expands via the FMS mechanism in leading order to the 'usual' $\pi^+ = |\bar{d}u\ra$.

This issue is even more important for a baryon. A baryon is a three quark state, as is necessary to obtain a strong singlet. But it is not possible to contract three weak fundamental charges to a singlet. Another fundamental charge is therefore necessary, which, e.g., can be obtained from another higgs. Thus, a gauge-invariant state would read, e.g., symbolically as either
\begin{align}
 \epsilon^{IJK} c_{ijkl} q_i^I q_j^J q_k^K h^{\dagger}_{\tilde{i}l}  \label{proton}
 \intertext{or}
 \epsilon^{IJK} q_i^I q_j^J q_k^K h^{\dagger}_{\tilde{i}i} h^{\dagger}_{\tilde{j}j} h^{\dagger}_{\tilde{k}k}
 \label{proton2}
\end{align}
\no depending whether the hadron has one or three open flavor indices, or to be more precise custodial indices, respectively. Capital indices $I,J,K$ are color indices, $i,j,k,l$ denote weak isospin indices and $\tilde{i},\tilde{j},\tilde{k}$ are custodial indices. The coefficient $c_{ijkl}$ for a conventional isospin-$1/2$ baryon, has to be chosen such that the resulting state is a gauge singlet regarding $\mathrm{SU}(2)$ and that the total wave function of the baryon is antisymmetric.

A straightforward example is the $\Delta^{++}$ resonance. Here, we have $\tilde{i}=\tilde{j}=\tilde{k}=1$ and the spin of all three quarks is aligned to form a totally antisymmetric wave function.

The situation is rather involved for the Proton. Choosing 
\begin{align}
 c_{ijkl} = a_1\, \epsilon_{ij}\delta_{kl} + a_2\, \epsilon_{ik}\delta_{jl} + a_3\, \epsilon_{jk}\delta_{il}
 \label{eq:c-proton}
\end{align}
with appropriate normalization factors, $a_1,a_2,a_3$, leads to a manifest gauge-invariant object which expands to the usual proton state for $\tilde{i}=1$ while $\tilde{i}=2$ would correspond to the field content of a neutron in leading order of the FMS expansion. In order to form a totally antisymmetrized wave function under the exchange of the three quarks, we have to assign the spin quantum numbers in an appropriate manner. This can be accomplished by using a spin wave function which is antisymmetric for the first two quarks, labeled by $i$ and $j$ for the first contribution on the right-hand side of Eq.~\eqref{eq:c-proton} and a symmetric spin wave function for the last two contributions which are symmetric under the exchange of $i$ and $j$ as usual.

Using the FMS mechanism, the Baryon states \eqref{proton} and \eqref{proton2} become again the ordinary baryon states, with the Baryon masses. Nonetheless, this implies that every Baryon, and most of the Mesons, have also at least one higgs component, though it may be small. They actually carry a custodial quantum number, rather than a flavor quantum number. In particular, as described in section \ref{s:gipt} for the Lepton case, this should influence the scattering of Hadrons, and should be detectable in principle. However, whether in practice this effect can be seen, given the QCD background, is currently speculative.

Of course, a bound state like a Lepton cannot decay in its constituent lepton and higgs, as this would yield a gauge non-invariant final state. It is only possible to decay into other bound states, e.\ g., a Muon into an Electron and accompanying Neutrinos, which recreate the known and measured \cite{pdg} decays.

All of this works because the higgs is a scalar particle. If it would have a different parity or spin, its presence in all of these bound states would change the spin and/or parity of the states, and thus would be an unambiguous signal of its presence. This is not the case, which is a strong conceptual hint that the higgs should be a scalar, as is consistent with experiment \cite{Aad:2015mxa,Aad:2013xqa}. Note also that the good agreement of perturbative results with experiments \cite{pdg} can be taken as an indication for the validity of the FMS mechanism. After all, it predicts perturbative results to be a good approximation, rather than to require non-perturbative methods. This is important, as the consideration of gauge invariance comes from a much deeper layer of the consistency of gauge theories.

\subsection{Generations as excitation spectra}

In the previous discussion it was essentially concluded that the concept of flavor has to be replaced by the custodial symmetry within a generation. Usually the flavors between different generations are considered to be something observable, and that there is a fundamental difference between, say, bottom and down. This is, however, already not the case in conventional perturbation theory. Also there the flavor identity is obtained from a combination of generation and weak isospin\footnote{For non-zero gauge coupling, the flavor of right-handed particles is actually an independent global symmetry. This should be carefully distinguished, but plays little role in the following. Note that we explicitly do not consider here the possibility of an unequal number of left-handed and right-handed generations, because we want to keep the anomaly cancellation within every generation as in the standard model.}. The difference in flavor comes entirely from the fact that the Yukawa interactions and the weak (isospin) interactions cannot be simultaneously diagonal in generation space, leading to the appearance of the CKM and PMNS matrices \cite{Bohm:2001yx}. Therefore already in conventional perturbation theory it is better to consider flavor not as an independent concept, but rather distinguish between intergeneration and intrageneration effects, and work with generations and doublets instead.

Let us combine this with the insights from the previous section that bound states, created from the interactions with the higgs field and/or the weak gauge bosons, are the physical degrees of freedom. Furthermore, these interactions are entirely responsible for all masses of the fermions\footnote{Neutrinos are here considered as ordinary Dirac neutrinos.}. This leads to an interesting speculation, which will be the subject of this subsection: Is it possible that the three generations are just internal excitations of a single generation? Then the standard model would be the effective theory of the excitation spectrum. While this may seem to be very far fetched at first glance, due to the large differences in scales, this looks at the second sight far less impossible: Even in the ordinary picture the bound states have huge mass defects. And the Yukawa interaction plays a dominant role.

In such a situation the left-handed bound states could have internal excitations, which would be described by further operator insertions. Correspondingly, the right-handed bare field operators would be supplemented by operators which have (weak gauge singlet) operator insertions. Simple versions of such operators can be constructed by adding gauge-invariant scalar operators, like $\phi^\dagger\phi$ operators. The excited states are thus similar to molecules, and the interaction is created by higgs exchange. This may seem to be quite impossible at first sight, since especially for light fermions the Yukawa interaction is very small. But already the lightest states, the ground states, obtain their entire mass by this interaction as a dynamical effect without gauge-fixing. The same is also true in a gauge with vanishing higgs vacuum expectation value \cite{Maas:2012ct}. This makes it much less unfeasible, once appreciated fully\footnote{Note that quantum mechanical arguments against such a possibility \cite{Grifols:1991gw} fail because the ground state already requires a relativistic treatment due to the large mass defects.}.

At any rate, the following part of this subsection is a gedankenexperiment. Let us check whether such a picture is compatible with present experimental knowledge.

If true, there needs to be a set of (at least) three mass eigenstates to be compatible with experiment. First of all, let us investigate the gauge invariant operator \eqref{fop} for one elementary isospin quark doublet $q$ in detail:
\begin{align}
 h^{\dagger}q = \begin{pmatrix} \phi^T \epsilon^T q \\ \phi^{\dagger}q \end{pmatrix}
 = \begin{pmatrix} \phi_2 u - \phi_1 d \\ \phi_2^* d + \phi_1^* u \end{pmatrix} \equiv \begin{pmatrix} U_o \\ D_o \end{pmatrix}
\end{align}
The field $q$ is the single elementary weak doublet of the theory. The (weak) gauge invariant-description of Up-type quarks and Down-type quarks are described by the gauge-invariant bound state operators $U_o = (\epsilon \phi^*)^{\dagger}q$ and $D_o = \phi^{\dagger} q$, analogously to the lepton case.

We would then interpret the ground state of the down-type quark operator $D_o=\phi^{\dagger} q$ as a physical $D=(\phi^{\dagger} q)_g$ quark.

In case that the operator $D_o$ has overlap with at least two higher excited states $S=(\phi^{\dagger} q)_*$ and $B=(\phi^{\dagger} q)_{**}$, we could interpret these states as Strange and Bottom quarks, respectively.
A similar consideration holds for the Up-type quarks. Therefore, the masses, and thus the Yukawa couplings, of the second and third generation would be a prediction of the theory, being the masses of the excited states. 

Furthermore, the elements of the CKM matrix could be predicted. For this, we have to switch to the mass eigenstates of the excitation spectrum of the operators $D$ and $U$. Let us now assume that the system is dominated by the lowest order operators in the Higgs field, namely $\phi^{\dagger} q$, $(\phi^\dagger \phi)\phi^{\dagger} q$, and $(\phi^\dagger \phi)^2 \phi^{\dagger} q$ for the Down-type quark for instance\footnote{Of course, any other operators with suitable quantum numbers would lead to qualitatively the same results, though with other quantitative overlaps with the different states.}. Each of these operators will have some overlap with the ground state $D$ as well as the excited states $S$ and $B$,
\begin{align}
 \phi^{\dagger} q &= \alpha_0\, D + \beta_0\, S + \gamma_0\, B + ... \, , \label{eq:down1} \\
 (\phi^{\dagger}\phi)\phi^{\dagger} q &= \alpha_1\, D + \beta_1\, S + \gamma_1\, B + ... \, , \label{eq:down2} \\
 (\phi^{\dagger}\phi)^2\phi^{\dagger} q &= \alpha_2\, D + \beta_2\, S + \gamma_2\, B + ... \, . \label{eq:down3}
\end{align}
In order to change the basis between two sets of operators, we can perform a unitary transformation $A_D$ for the Down-type quarks and similar for the Up-type quarks with a unitary matrix $A_U$. Investigating transitions between two custodial eigenstates in the basis of the mass eigenstates involves elements of a unitary matrix $V=A_U^\dagger A_D$. But this is exactly the same situation as already present in the standard model itself \cite{Bohm:2001yx}, just that here the bound states are replaced by the elementary fermions, and the rotation is obtained from the CKM matrix in the quark sector and PMNS matrix in the lepton sector. In total analogy to the standard model case, the unitary matrix $V$ has nine free parameters from which we can remove five by appropriate phase rotations of the excited and ground states while a global phase factor is redundant. Thus, we remain with three rotation angels and an additional phase which characterizes CP violation, all of which would be computable from the coefficients on the right-hand side of Eqs.~\eqref{eq:down1}-\eqref{eq:down3} and similar equations for the Up-type quarks.

In this gedankenexperiment the perturbative treatment of the three-generation standard-model is then just a low-energy effective theory for the excitation spectrum interactions, similar to chiral perturbation theory. The actual underlying theory is a one-generation standard model, where the generations are created as a dynamical bound state effect. The masses of the excitations in the effective theory are parameters obtained from experiment. Likewise, the CKM/PMNS elements are just the decay matrix elements of excited states to less excited states or to the ground states\footnote{Again, none of the states can decay in its constituents, as this would not be a gauge-invariant process.}. Just as in chiral perturbation theory, these are put in the three-generation low-energy effective Lagrangian by measurement, while they could be calculated in the underlying one-generation Lagrangian. Only four parameters remain in the one-generation Lagrangian: The four (two quark and two lepton) Yukawa couplings to the higgs.

This would reduce the number of free parameters substantially. Note that even CP violation can be incorporated into this picture as a dynamical mixing effect. For instance, the Kaon decay translates in a straightforward manner for the excited states.

The biggest theoretical challenge here is the necessity to show that such an excitation spectrum exists. While technically demanding, and practically yet out of reach, this is in principle possible. A confirmation would require not only the existence of the spectrum, but is highly constrained by a multitude of high-precision tests \cite{pdg}, e.g., the bounds on flavor changing neutral currents. The alteration in the number of elementary degrees of freedom would probably also have cosmological implications.

The biggest experimental challenge yet is probably the measurement of the number of (light) generations from the deacys of the $Z$ \cite{pdg}, which yields three. However, this result is not in contradiction, when following the order of theories. Only at energies of the order or larger than the electroweak scale/higgs mass the one-generation structure appears. In the low-energy effective theory all parameters of the standard-model are fixed by measurements, which, of course, include the bound-state effects. Any further imprints of the bound-state structure can only appear as higher-dimensional operators, which, however, will include the higgs field as the relevant degree of freedom. Thus, these corrections are at least as suppressed as other higgs effects. Thus, below the electroweak/higgs scale, and especially at the $Z$ mass, probing the inner structure is suppressed, and dominated by the perturbative corrections of the three-generation standard model. Thus, only the bound states are apparent, but not their structure.

Still, this requires that there exists no fourth light excitation in the Lepton sector, like a fourth would-be Neutrino. Whether this is indeed the case returns to the requirement of an explicit calculation to answer the question: How many excited states are there? And what masses do they have? It is experimentally consistent that there are only three light ones, but an explicit theoretical calculation is necessary.

However, at least the considerations above show that a mapping from the standard model parameters to the parameters of the excitation spectrum is possible. Whether the parameters of the excitation spectrum indeed fit the standard model parameters has to be shown by a detailed analysis of the excitations. At the current time this is not yet possible because no genuine non-perturbative method is yet practically able to cope with the huge differences between the levels, especially as these states are unstable. 

But, even if no such excitation spectrum exists, the electroweak sector is still described in a gauge invariant manner for three generations of elementary fermions with the aid of the FMS mechanism for each of the generations separately.

There is a possibility to test experimentally the implications of both the FMS mechanism in the fermion sector and this additional speculation. If these are bound states, they have an inner structure, which should be possible to probe. In analogy to other bound states, it is expected that the inner structure becomes visible if the involved energy scales are of the order of the masses of the constituents and/or mass defects. In the present case, this is of order the higgs mass. Thus, this requires to probe the inner structure with energies of order the higgs mass.

The remainder of this work is dedicated to estimate how the response of such a bound state to a probing could possibly look like, and where it would possibly be worthwhile to search for it. Lacking a possibility to determine the inner structure in case of the above discussed speculation, we do this by considering again the three-generations case.

\section{Gauge-invariant perturbation theory}\label{s:gipt}

\subsection{Fundamental expressions}

To this end, it is necessary to consider a realistic probing experiment. Though protons have also such an inner structure, as the operator \pref{proton}  suggests, it appears at first sight unlikely that this structure would be easy to isolate in proton-proton collisions like at the LHC, due to the QCD background. We therefore turn to lepton collisions, especially Fermion pair production in $E^+ E^-$ collisions. In the FMS picture this is a collision of electron-higgs bound states. Here, we concentrate on Muons, Bottom quarks, or Top quarks in the final state, denoted collectively as $F$.

The corresponding gauge-invariant matrix element is given by
\be
{\cal M}=\la {\cal O}^{\nu e}_2(p_1){\bar{\cal O}}^{\nu e}_2(p_2){\cal O}_F(q_1){\bar{\cal O}}_F(q_2)\ra\label{fullm}
\ee
\no in the center of mass system $\vec p_1+\vec p_2=\vec q_1+\vec q_2=0$ and $(p_1+p_2)^2=s$. Only the lower-component custodial states appear in the initial state as we are interested in $E^- E^+$ scattering. (For the $M^- M^+$ final state, we have to investigate the second component of the custodial doublet operator for the second generation of leptons while for the case of $B\bar{B}$ or $T\bar{T}$ we have to use the second or first component of $\mathcal{O}^{tb}$, respectively). We neglect that quarks would carry a color charge and would hadronize, assuming this will happen on a sufficiently long time-scale to not affect any of the investigations here. This is motivated by the fact that the QCD scale is much smaller than the electroweak scale.

Following the rules of gauge-invariant perturbation theory in section \ref{s:custodial}, the gauge-invariant operators have to be rewritten in a fixed gauge. We choose for this a 't Hooft gauge \cite{Bohm:2001yx}. Since at tree-level the gauge-parameter dependence explicitly cancels according to the Goldstone boson equivalence theorem, which we checked, we do not specify the gauge further.

To identify what are the dominant subprocesses, we order the expression by the powers of vacuum expectation values. To avoid a unnecessary lengthy expression, the result is given here only symbolically for some of the relevant terms,
\begin{align}
\begin{split}
{\cal M} &\approx v^4\la e^+e^-\bar{f}f\ra+v^2\la\eta^\dagger\eta e^+e^-\bar{f}f\ra \\
&\quad +\la\eta^\dagger\eta\eta^\dagger\eta e^+e^-\bar{f}f\ra + \text{rest}
\label{scatt1}.
\end{split}
\end{align}
\no The first expression is the usual matrix element, while the other two involve additional higgs particles $\eta$ (the fluctuation field proportional to the direction of the vacuum expectation value in a 't Hooft gauge) in the initial or final states. Note that only BRST singlets can appear here as initial and final states, and correlation functions involving, e.g., goldstones will vanish.

There are also many other contributions with an odd number of higgs particles. These essentially exchange a higgs particle with the vacuum or one of the fermions. However, we neglect initial state radiation of higgs particles and require an exclusive measurement of the final state. Thus, such correlation functions have the wrong number of legs, and will therefore not contribute.

The explicitly non-trivial two types of expectation values, i.e., order $v^2$ and 1, in Eq.~\pref{scatt1} expand at tree-level as
\begin{align}
{\cal M} &\approx 
v^2 \Big( \la\eta^\dagger\eta\ra\la e^+e^-\bar{f}f\ra  +  \la e^+e^-\ra\la\eta^\dagger\eta\bar{f}f\ra  \notag\\ 
&\qquad\quad +  \la\bar{f}f\ra\la e^+e^-\eta^\dagger\eta\ra\Big) \label{eq:scatt2} \\
&\quad +\la\eta^\dagger\eta\eta^\dagger\eta\ra\la e^+e^-\bar{f}f\ra+\la e^+e^-\eta\eta^\dagger\ra\la\eta^\dagger\eta \bar{f}f\ra + ... \, . \notag
\end{align}
\no In the first three contributions ${\sim}\,v^2$ always a single-particle propagator appears, which is part of the initial or final state. It does not interact at this level, and thus corresponds to the spectators of the initial or final state, but it has a vacuum insertion. This can be interpreted as that the spectators are absorbed or emitted by the vacuum, and thus, the corresponding quantum numbers are absorbed or emitted from the condensate. Alternatively, as in the QCD case, this can be interpreted to be the remnants moving into the beam-pipe or as the electroweak equivalent of the hadronization process. At any rate, as spectators they will not play a role in the following, and can be thought to be absorbed in the PDFs or fragmentation functions as in the QCD case to be introduced below in section \ref{ss:pdfs}.

The first term in Eq.~\eqref{eq:scatt2} modifies the leading perturbative expression by a certain factor\footnote{Note that this is now a double expansion in both the couplings and $\eta$. At any given order parameters like $v$ have to be fixed again as usual such that the results agree with experiment \cite{Bohm:2001yx}.}. The second expression is the really interesting one: Here, the electron and positron of the initial state are spectators, and the generation of the final state fermions is entirely due to the interactions of the higgs constituents. This contributes at the same order in the perturbative expansion. The last of the three terms is dominated by the higgs-electron Yukawa coupling, and will therefore be ignored, as it is quantitatively irrelevant.

The two contributions in the third line of Eq.~\eqref{eq:scatt2} describe individual scatterings of the constituents of the bound states, rather than the scattering of the full bound states. This later set is of higher order in a perturbative expansion than the leading order, and can at leading-order be ignored.

Also, there are always more versions of all of these expressions, depending on the actual arguments of the higgs fields, as they can either propagate or act as spectators. Moreover, there is also momentum partitioning involved.

The question remains how to account for these two effects, the modification of the original amplitude of the leading order in \pref{scatt1}, and the appearance of the second process.

\subsection{A PDF-type language}\label{ss:pdfs}

Inspired by how this is done in QCD \cite{Bohm:2001yx}, we will use a PDF-type language in the following.

The higgs sector is not asymptotically free, but this takes only effect at energies many orders of magnitude larger than considered here. This should therefore be of minor concern.

Furthermore, the large mass of the higgs, providing a hard scale in comparison to the initial bound state mass, and the general weak coupling suggest that factorization is viable in the present case. However, a more fundamental investigation will be required eventually. Especially, for non-generalized PDFs it is usually required that all partons are essentially massless, i.\ e., $m_\mathrm{parton}^2\ll s$. While this does not hold for the Higgs at the low-energy versions of the planned colliders, the Applequist-Carrazone theorem \cite{BeiglboCk:2006lfa} suggests that this will yield corrections in powers of $m_h^2/s$, which will be neglected here for simplicity. Alternatively, if the reader wishes, the results can be easily extrapolated to $s\gg m_h^2$, where the realm of ordinary PDFs is reached, and factorization should work as usual. Best would be, of course, to use generalized PDFs \cite{Lorce:2013pza,Diehl:2015uka}, but this is a demanding future research project.

We will also assume that the final state fragmentation of the elementary fermion pair $f$ and $\bar{f}$ will not substantially affect the process itself, and therefore do not include any fragmentation functions. This is equivalent to assuming that the hard produced elementary fermions hadronize to 100\% into the corresponding bound states.

To this end, we start again from \pref{fullm}, and rewrite the corresponding cross-section using PDFs as \cite{Bohm:2001yx}
\begin{align}
&\sigma_{E^+E^-\to\bar{F}F}(s) \label{pdf} \\
&=\sum_i\int_0^1 dx\int_0^1 dy f_i(x) f_i(y)\sigma_{\bar{i}i\to\bar{f}f}(xp_1,yp_2).\notag
\end{align}
\no where the subscript $i$ denotes the involved elementary particles in the initial state.

\begin{figure*}[ht]
\vspace{1.5cm}
\begin{fmffile}{electron}
\begin{fmfgraph*}(70,30)
\fmfleft{i1,i2}
\fmfright{o1,o2}
\fmflabel{$e^+$}{i1}
\fmflabel{$e^-$}{i2}
\fmflabel{$\bar{f}$}{o1}
\fmflabel{$f$}{o2}
\fmf{fermion}{i1,v1,i2}
\fmf{fermion}{o1,v2,o2}
\fmf{photon,label=$\gamma$}{v1,v2}
\end{fmfgraph*}
\hspace{0.5cm}
\begin{fmfgraph*}(70,30)
\fmfleft{i1,i2}
\fmfright{o1,o2}
\fmflabel{$e^+$}{i1}
\fmflabel{$e^-$}{i2}
\fmflabel{$\bar{f}$}{o1}
\fmflabel{$f$}{o2}
\fmf{fermion}{i1,v1,i2}
\fmf{fermion}{o1,v2,o2}
\fmf{photon,label=$Z^0$}{v1,v2}
\end{fmfgraph*}
\hspace{0.5cm}
\begin{fmfgraph*}(70,30)
\fmfleft{i1,i2}
\fmfright{o1,o2}
\fmflabel{$e^+$}{i1}
\fmflabel{$e^-$}{i2}
\fmflabel{$\bar{f}$}{o1}
\fmflabel{$f$}{o2}
\fmf{fermion}{i1,v1,i2}
\fmf{fermion}{o1,v2,o2}
\fmf{dashes_arrow,label=$\eta$}{v1,v2}
\end{fmfgraph*}
\hspace{0.5cm}
\begin{fmfgraph*}(70,30)
\fmfleft{i1,i2}
\fmfright{o1,o2}
\fmflabel{$e^+$}{i1}
\fmflabel{$e^-$}{i2}
\fmflabel{$\bar{f}$}{o1}
\fmflabel{$f$}{o2}
\fmf{fermion}{i1,v1,i2}
\fmf{fermion}{o1,v2,o2}
\fmf{dashes_arrow,label=$\chi$}{v1,v2}
\end{fmfgraph*}
\end{fmffile}
\\[3mm]
\rule{\textwidth}{0.2pt}
\\[7mm]
\centering
\begin{fmffile}{higgs}
\begin{fmfgraph*}(70,30)
\fmfleft{i1,i2}
\fmfright{o1,o2}
\fmflabel{$\eta$}{i1}
\fmflabel{$\eta$}{i2}
\fmflabel{$\bar{f}$}{o1}
\fmflabel{$f$}{o2}
\fmf{dashes_arrow}{i1,v1,i2}
\fmf{fermion}{o1,v2,o2}
\fmf{dashes_arrow,label=$\eta$}{v1,v2}
\end{fmfgraph*}
\hspace{0.5cm}
\begin{fmfgraph*}(70,30)
\fmfleft{i1,i2}
\fmfright{o1,o2}
\fmflabel{$\eta$}{i1}
\fmflabel{$\eta$}{i2}
\fmflabel{$\bar{f}$}{o1}
\fmflabel{$f$}{o2}
\fmf{dashes_arrow}{i1,v1}
\fmf{dashes_arrow}{i2,v2}
\fmf{fermion}{o1,v1,v2,o2}
\end{fmfgraph*}
\hspace{0.5cm}
\begin{fmfgraph*}(70,30)
\fmfleft{i1,i2}
\fmfright{o1,o2}
\fmflabel{$\eta$}{i1}
\fmflabel{$\eta$}{i2}
\fmflabel{$\bar{f}$}{o1}
\fmflabel{$f$}{o2}
\fmf{dashes_arrow}{i1,v1}
\fmf{phantom}{v1,o1} 
\fmf{dashes_arrow}{i2,v2}
\fmf{phantom}{v2,o2} 
\fmf{fermion}{v1,v2}
\fmf{fermion,tension=0}{v1,o2}
\fmf{fermion,tension=0}{o1,v2}
\end{fmfgraph*}
\end{fmffile}
\caption{\label{feynman}Feynman diagrams of the relevant processes. The top line gives the contributions for $\sigma_{\bar{e}e\to\bar{f}f}(xp_1,yp_2)$ and the bottom line for $\sigma_{\bar{\eta}\eta\to\bar{f}f}(xp_1,yp_2)$. The Higgs components $\eta$ and $\chi$ are the fluctuation and uncharged Goldstone fields of the Higgs, respectively.}
\end{figure*}
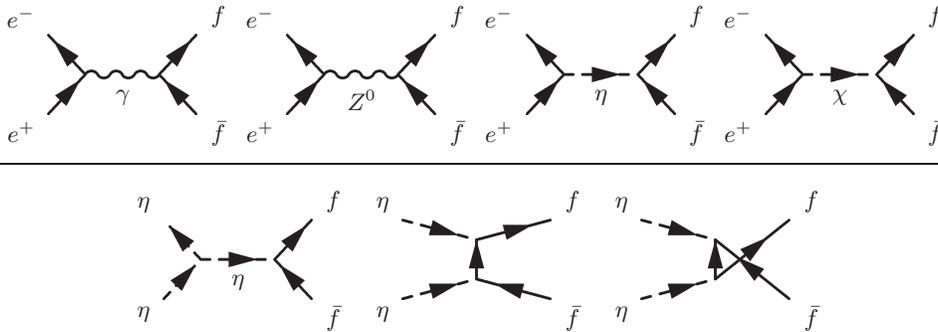

The cross-sections $\sigma_{\bar{i}i\to\bar{f}f}$ are the elementary perturbative cross-sections for the corresponding processes, which will be evaluated to lowest non-vanishing order in all couplings. We neglect all transverse momentum components, as the higgs mass is assumed to be sufficiently large compared to them. We consider here the cases of LEP(2), CEPC, and the ILC, i.e., up to an energy of $\sqrt{s}=1$ TeV. However, the results can be extrapolated straightforwardly to any arbitrary $s$. The calculations of $\sigma_{\bar{i}i\to\bar{f}f}(xp_1,yp_2)$ is a straightforward application of standard perturbation theory \cite{Bohm:2001yx}, which we will not detail here. The explicit expressions can be found in \cite{Egger:2016mt}. The Feynman diagrams included are shown in figure \ref{feynman}.

Because the PDF language requires the partons to be on-shell, no effect of the bound state structure can arise in the present approximation for $s<4m_h^2$. Therefore, in this energy range just the electrons and positrons in the initial state contribute to the cross-section. This can also be seen from the fact that the corresponding elementary cross-section $\sigma_{\eta\eta\to\bar{f}f}$ vanishes below this threshold. This can be made explicit by rewriting \pref{pdf} as
\begin{align}
&\sigma_{E^+E^-\to\bar{F}F}(s) \label{pdf2} \\
&=\theta(4m_h^2-s)\sigma_{e^+e^-\to\bar{f}f}+\theta(s-4m_h^2)\times\notag\\
&\times\sum_i\int_0^1 dx\int_0^1 dy f_i(x) f_i(y)\sigma_{\bar{i}i\to\bar{f}f}(xp_1,yp_2).\notag
\end{align}
Thus, for center-of-mass energies below this threshold, no consequences of the bound state structure of the initial state can be felt. This includes the LEP(2) energies. Relaxing this on-shell condition requires not only virtual particles as initial states in the hard processes, but the use of off-shell PDFs \cite{Lorce:2013pza,Diehl:2015uka}. The complexity of such an endeavor is far beyond the scope of the present investigation, and therefore also left for future research. Still, off-shell contributions are at leading order usually suppressed compared to on-shell contributions. Even the largest LEP2 energies would require the higgs particles to be very far off-shell. This gives an additional reason why none of the bound state structure has been seen at LEP2 energies.

Based on the considerations derived from \pref{scatt1} and \pref{eq:scatt2} the sum in \pref{pdf2} should run over electrons/positrons and the higgs particles. Then $f_i$ is one of the two PDFs for the electron and the fluctuation field of the higgs, $f_e$ and $f_\eta$, respectively. 

The PDFs have to fulfill the sum rules
\bea
\int \!\! dx \, f_e&=&1\label{sumrulec}, \\
\int \!\! dx \, x(f_e+f_\eta)&=&1\label{sumrulee},
\eea
\no encoding the total charge and the total energy, respectively.

The next complication arises from the $s$-channel exchange in this process. Because the tree-level propagators have singularities at their masses, the integrals in \pref{pdf2} over $x$ and $y$ are not well-defined. To avoid this, the propagators for the $Z$ and the $\eta$, and for the $\chi$ to keep gauge invariance in the interplay with the $Z$ propagator, are given their physical width by replacing their singularities as
\be
\frac{1}{p^2-m^2}\to\frac{1}{p^2-M^2+iM\Gamma}\nn
\ee
\no where $\Gamma$ is their observed width \cite{pdg}. As these widths are in all cases rather small, this does not substantially affect the results even close to the pole. Note that we include the electron mass in the calculations, and therefore no treatment of the photon pole is needed.

\subsection{Pair production process at the ILC and CEPC}\label{s:dy}

The last step is to specify the PDFs in \pref{pdf2}. Given that in the present on-shell formulation no experimental data can be used to constrain the PDFs yet -- this will require either CEPC or ILC -- here ans\"atze will be made.

As the bound states are two-particle states we make the ansatz that the momentum should be either distributed such that one particle carries most of the momenta and the other almost none, or both could carry about the same amount. A suitable structure is
\begin{align}
\begin{split}
f_i(x)= \frac{1}{\sqrt{2\pi w^2}}  \Big( & ae^{-\frac{x^2}{2w^2}}  +  be^{-\frac{(x-1/2)^2}{2w^2}}+ce^{-\frac{(x-1)^2}{2w^2}}\Big).
\end{split}          
\label{pdfnt}
\end{align}
\no Note that if the energy available to a parton is smaller than $m_i^2$ the energy fraction is insufficient to put the parton on-shell. Then the cross-sections involving this parton vanish.

In principle, each of the peaks could have its own width, but for the present exploratory study this provides too much freedom. Thus, the three prefactors are mainly parameterizing the relative importance of the three structures. Out of the eight free parameters the two sum rules \prefr{sumrulec}{sumrulee} constrain two.

The case of ordinary perturbation theory is recovered for
\bea
f_e(x)&=&\delta(x-1)\label{pdft1}\\
f_\eta(x)&=&0\label{pdft2},
\eea
\no which automatically fulfills the sum rules. This is also the form which arises if $s<(2m_h)^2$, because then the higgs constituents cannot come on-shell. Therefore, nothing changes below the two-Higgs threshold.

Now the following situation arises. Because the cross-section $\sigma_{e^+e^-\to\bar{f}f}$ rises quickly with decreasing $s$, any substantial contribution at $x<1$ immediately increases the total cross-section substantially. Thus, any contribution for $b_e$ is problematic. This is even more so if $w\gtrsim xm_e/s$, as then soft electrons will even more enhance the effect due to photon-exchange. Thus, if the cross-section should not become much larger, this requires $f_e$ to be strongly peaked around $x=1$, i.\ e.\ having a very small width $w$. The prefactor is then entirely determined by the charge sum-rule \pref{sumrulec}. However, the energy sum-rule \pref{sumrulee} then allows only a negligible contribution around $x=1$ for $f_\eta$, as long as $a_\eta$ is not large and negative. The only way out is having $a_e$ to be non-zero, and comparatively large, at small width.

Pushing these considerations to the extreme yields an ansatz
\bea
f_e(x)&=&a_e\delta(x)+c_e\delta(x-1)\nn\\
f_\eta(x)&=&a_\eta\delta(x)+c_\eta\delta(x-1)\nn.
\eea
\no The sum-rules yield $a_e+c_e=1$ and $c_e+c_\eta=1$, and the value $a_\eta$ is undetermined. The expression \pref{pdf} can then be analytically calculated. Normalizing the results to the conventional perturbative results yields
\bea
\frac{\sigma_{E^+E^-\to\bar{F}F}}{\sigma_{e^+e^-\to{\bar f}f}}&=&\frac{c_e^2\sigma_{e^+e^-\to{\bar f}f}+(1-c_e)^2\sigma_{\eta^\dagger\eta\to{\bar f}f}}{\sigma_{e^+e^-\to{\bar f}f}}\nn\\
&=&c_e^2+(1-c_e)^2\frac{\sigma_{\eta^\dagger\eta\to{\bar f}f}}{\sigma_{e^+e^-\to{\bar f}f}}\label{ratio}
\eea
\no Any softening of the $\delta$-functions to the exponentials of \pref{pdfnt} will diminish the effect, as long as $a_\eta\ge 0$ is required. As in the present tree-level ansatz the PDF still have a probability interpretation \cite{BeiglboCk:2006lfa}, we will refrain from allowing a negative $a_\eta$. For the same reason, this requires $c_e\le 1$.

For light final states the cross-section $\sigma_{\eta^\dagger\eta\to\bar{f}f}$ is negligible. Therefore, in these cases the ratio will essentially drop to $c_e^2$ above the two-higgs threshold. Hence, only for the top a substantially other behavior can be expected for $c_e\lesssim 1$.

This establishes the qualitative features of our results. A quantitative statement requires knowledge of the internal structure, which in the form \pref{ratio} is completely encoded in the value of $c_e$.

\begin{figure*}
\includegraphics[width=0.5\linewidth]{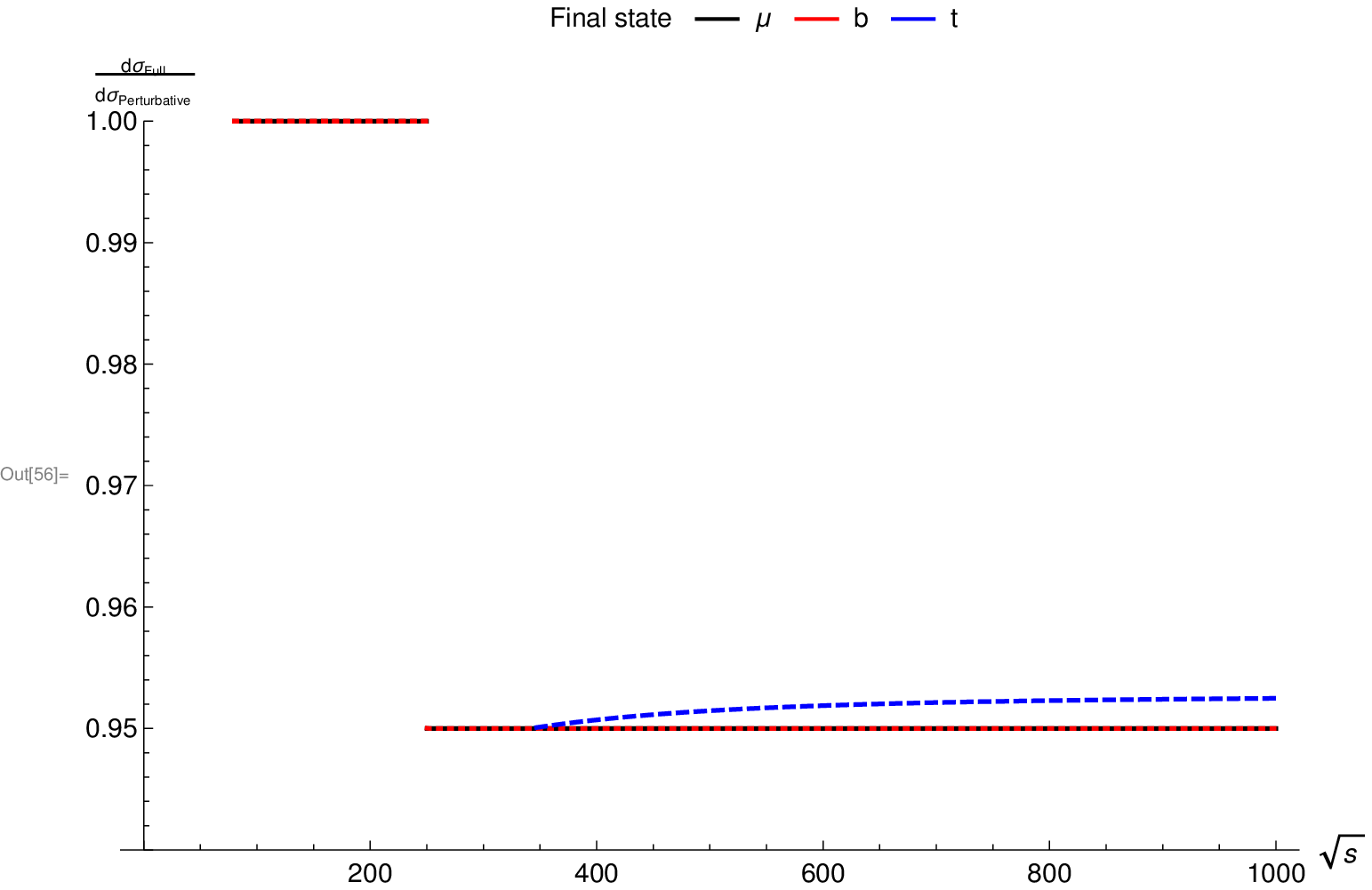}\includegraphics[width=0.5\linewidth]{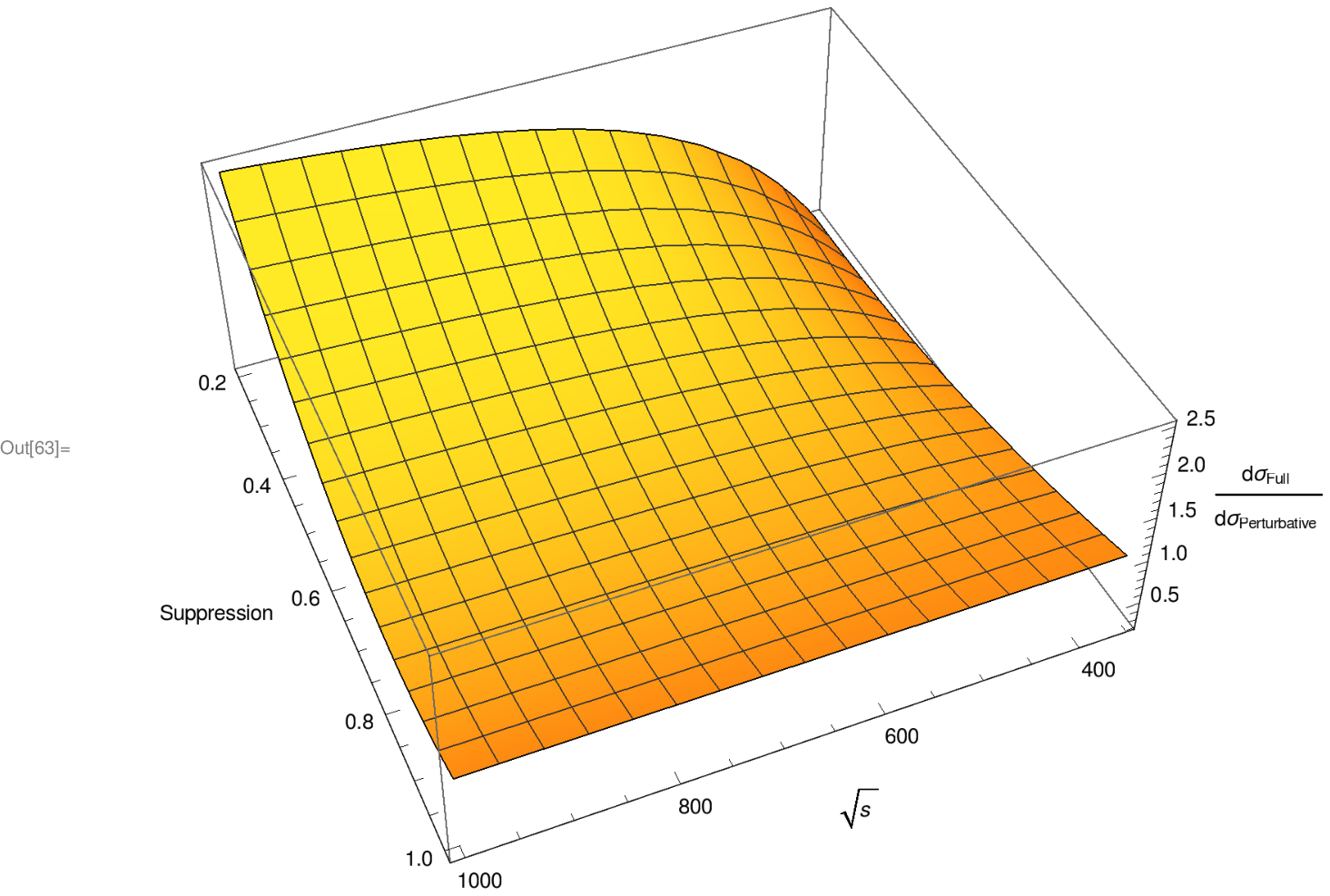}
\caption{\label{res}The left-hand side shows the ratio of the full case to the trivial case \pref{ratio} at rapidity zero and as a function of $\sqrt{s}$ in GeV for different final states for $c_e^2=0.95$. The right-hand-side shows the same ratio for the top final state as a function of both $a_e^2$ and $s$. Note that here the differential cross-section in the rapidity are shown, rather than the integrated one, to avoid problems due to collinear singularities. However, \pref{ratio} remains unchanged when the total cross-sections are replaced by the differential cross-sections.}
\end{figure*}

Assuming values for $c_e$, quantitative examples of the effects are shown in figure \ref{res}. It shows all the qualitative features, but the most dominant effect is the quick decay with $s$ to the value determined by $c_e^2$, except for the top. Thus, the depletion of the cross-section above the two Higgs threshold would be the signature for the substructure, if the qualitative features of the approximations performed here carry over to the full case. Note that since both cross-sections depend, up to a prefactor, asymptotically in the same way on $s$ the signature remains also at higher energies, especially at energies where it would be safe to neglect the mass of the higgs.

Taking the result literally, and assuming that $c_e$ is not too different from one, the picture of a bound state emerges in which most of the energy is carried by the electron, while the higgs component carries essentially almost nothing of the energy. Zero momentum is, however, equivalent to no strong localization inside the bound state, and the higgs field fills out the bound state more or less homogeneously. This is a very similar picture to the usual one of a constant higgs vacuum expectation value.

Whether this is indeed observable at CEPC or ILC now depends entirely on the actual size of $c_e$, or the actual shape of the PDFs. But it is a statement of where to look for the effect experimentally, and it seems to be in range for these experiments, provided all assumptions made are sufficiently good. 

While the quantitative values depend on many more details, a modification is the unambiguous prediction of the theory. The only question is how large, and here the parameters of the standard-model very likely will require a high precision measurement.

There are several future avenues how to obtain more quantitative predictions theoretically.

To incorporate insights from existing data would require to extend the PDFs such that also off-shell initial states are included. Though this is possible in principle \cite{Lorce:2013pza,Diehl:2015uka}, this will not be an entirely trivial exercise. Especially, this raises the question of whether other off-shell (sea) degrees of freedom, like $w$ and $z$ bosons and the would-be goldstone bosons, should not also be included. Further charged particle PDFs would reduce the importance of the charge sum rule \pref{sumrulec}, which currently strongly restricts the shape of the electron PDF.

Another possibility would be to use lattice calculations or functional methods to determine the PDFs, at least for the quenched case. While in principle this may be possible \cite{Nguyen:2011jy,Lin:2014zya,Chen:2016utp}, this is yet even for QCD a pioneering area. Nonetheless, as there are no detected substantial deviations in available experimental results \cite{pdg}, current experiments can only serve as constraints to the PDFs. Thus, such non-perturbative investigations will probably be necessary to make precise quantitative predictions for the CEPC and the ILC.

\section{Summary}\label{s:sum}

We have discussed the implications of gauge-invariant perturbation theory, based on gauge invariance and the FMS mechanism \cite{Frohlich:1980gj,Frohlich:1981yi}, for the flavor sector of the standard model. In particular, we have analyzed how the necessary compositness of gauge-invariant states describing the standard model fermions requires to replace the usual flavor quantum number by a custodial quantum number. We have also investigated whether it is possible that the standard model's generation structure could be dynamically generated.

In the second part of this work we turned to experimental consequences of these considerations. We investigated pair production of fermions at the CEPC and the ILC in gauge-invariant perturbation theory. To account for non-trivial effects, we transferred the PDF language of QCD bound-state scattering to the present case. We found that the substructure will play a role above the two-Higgs threshold, but it is most likely a tiny effect. This suggests to study the corresponding process at very high precision, but this is fortunately anyway a primary goal of these experiments. We also found a rather similar behavior for all final states, except the top.

In absence of data for this case, and first-principle non-perturbative calculations of the PDFs even in QCD still being challenging, we can only conclude that there is a potential for deviations from the standard perturbative picture. This requires a thorough understanding, as any such deviations could be mistaken for new physics effects, while being just ordinary standard-model physics. As noted in section \ref{s:custodial}, there may even be a potential for deviations in hadron-hadron collisions in high precision data, like at the high-luminosity LHC or the future FCC. However, in this case the situation is much more involved than in the $e^+e^-$ case.\\

\no{\bf Acknowledgments}\\

\no We are grateful to L.\ Pedro and P.\ T\"orek for helpful discussions. Moreover, we are indebted to H. Gies for valuable discussions and a critical proof reading of the manuscript. R.\ S.\ acknowledges support by the Carl-Zeiss foundation.

\appendix

\bibliographystyle{bibstyle}
\bibliography{bib}


\end{document}